# Stable room temperature magnetic graphite

Helena Pardo[#], Fernando M. Araújo-Moreira[§], Ricardo Faccio[#], Oscar F. de Lima[¥], Alexandre J. C. Lanfredi[§], Cláudio A. Cardoso[§], Edson R. Leite[‡], Giovani Zanelatto[§] & Álvaro W. Mombrú[#]

[#] *Crystallography, Solid State and Materials Laboratory (Cryssmat-Lab), Dequifim, Facultad de Química, Universidad de la República, P.O. Box 1157, Montevideo, URUGUAY*

[§] *Grupo de Materiais e Dispositivos-CMDMC, Departamento de Física e Engenharia Física, UFSCar, Caixa Postal 676, 13565-905, São Carlos SP, BRAZIL*

[¥] *Instituto de Física "Gleb Wataghin", UNICAMP, 13083-970, Campinas SP, BRAZIL*

[‡] *Laboratório Interdisciplinar de Eletroquímica e Cerâmica-CMDMC, Departamento de Química, , UFSCar, Caixa Postal 676, 13565-905, São Carlos SP, BRAZIL*

**Carbon materials are attracting increasing attention due to the novelty of the associated physical properties and the potential applications in *high-tech* devices. The possibility to achieve outstanding properties in macroscopic carbon materials opens up a profusion of new striking applications. Magnetic properties induced by defects on graphite structures, such as pores, edges of the planes and topological defects, have been theoretically predicted. The possible coexistence of $sp^3$ and $sp^2$ bonds have been also invoked to predict this behavior (for a review, see ref. 1). Some reports have proved the existence of weak ferromagnetic-like magnetization loops in highly-oriented pyrolytic graphite (HOPG) (ref. 2-3). Very recently two reports showed that the existence of ferromagnetism in pure carbon is unambiguously possible (ref. 4-5). Here we report on a novel and inexpensive chemical route consistent in a controlled etching on the graphite structure to obtain macroscopic amounts of magnetic pure graphite. This material has a strong magnetic response even at room temperature where it can be attracted by a commercial magnet and would be the experimental confirmation for the defect induced magnetism previously predicted.**

The chemically modified graphite here reported was produced by a vapor phase redox reaction in closed nitrogen atmosphere ($N_2$, 1 atm.) with copper oxide (CuO). A few grams of both powders, CuO (Merck, ppa) and graphite (Fluka, CAS Number 7782-42-5, lot 426277/1, granularity < 0.1 mm) were placed at different alumina crucibles in a sealed atmosphere, inside a tube furnace. The reaction took place at 1200ºC, during 24 hours. After the reaction was finished, the CuO was partially reduced to Cu(0) and, in the other container, the carbon material showed a decrease in volume, specially in the side closest to the CuO crucible. Two clearly different regions could be observed: an upper layer, black and opaque, with amorphous aspect, and a lower layer, more crystalline than the original pure graphite. The material from the upper layer was the one showing the magnetic behavior, detectable up to room temperature. The separation of both materials was carefully achieved with the aid of a magnet.

This magnetic graphite was produced by the reaction of pristine graphite with controlled amounts of oxygen released from the decomposition of CuO at high temperature. This chemical attack created pores and stacking structures and increased the exposed edges of the graphene planes, producing a foamy-like graphite. The removed carbon may also recrystallize on the graphite crystallites surfaces at the same crucible, thus creating the so-called lower region. This vapor phase reaction took place in an inert gas environment. In this case both $N_2$ and Ar have been used with similar results, discarding the specific role of $N_2$ as a reagent or catalyst. The reproducibility of the method and purity of the materials, with very special concern on the presence of metallic impurities, were regarded. Several samples have been prepared and all of them exhibited the magnetic behavior here described. Since the presence of any kind of ferromagnetic impurity must be avoided, we have carefully determined the chemical purity of the samples with AAS (atomic absorption spectroscopy) using a Shimadzu AA6800 spectrometer and checked these results with XRF (x-ray fluorescence analysis) and EDS (energy dispersive spectroscopy), comparing the results obtained for the pristine and the modified graphite. The total content of iron in the samples was in the 40-60 ppm range. The results did not indicate any increase of metallic impurities with respect to the original pristine graphite. If these impurities would have been the cause of the magnetic effect here reported, both graphite samples, the modified and the pristine ones, should exhibit the same magnetic behavior, which clearly is not the case.

We have studied the obtained magnetic graphite samples by scanning electron microscopy (SEM), using a Jeol JSM 5900LV microscope. Raman spectra were recorded with a J-Y T64000 triple (subtractive) Micro-Raman spectrometer, using a 50x, 0.5n.a. objective, to focus 3mW of the 514.5 nm line of an Ar+ laser (laser spot put over a similar graphite crystal in both samples, the pristine and the modified graphite, far from the crystal border). We also used magnetic force microscopy (MFM), and magnetometry as a function of temperature and applied magnetic field, to magnetically characterize the samples. MFM measurements were performed using a Nanoscope III, by Digital Instruments. Magnetic measurements were performed using a MPMS-5T Quantum Design magnetometer.

SEM micrographs have given enough evidences about the topological changes occurring at the micro-structural scale. Figure 1a. shows a micrograph from the pristine graphite powder used as reagent. Figure 1b. shows a graphite grain after chemical attack, exhibiting several pores distributed all over the sample. It also can be observed that a large dispersion of pore sizes can be achieved, with diameters ranging from a few nanometers to approximately 1 μm. Figure 1c. clearly shows a pore having about 1μm diameter that goes through many graphene layers. In some regions where the etching process affected the boundaries of the graphene layers, other complex microstructures have been observed, like the stacking structure shown in Figure 1d.. Micro Raman analysis (see Figure 2) of the pristine graphite exhibited a peak at 1580 cm$^{-1}$, showing the good quality of the reagent used in this procedure. The modified magnetic graphite clearly showed the occurrence of the 1350 cm$^{-1}$ peak, well-known as the "disordered" D-band. A broadening of the 1580 cm$^{-1}$ peak was also observed. Both signals were indicative that the crystal could no longer be regarded as infinite, and that the presence of defects was clearly significant. The presence of magnetic regions at the micro-scale has been verified by MFM. Also, the role of the pores and the micro-structural new features were studied simultaneously with atomic force (AFM) and magnetic force microscopy (MFM), trying to relate the topography to the magnetic signal. Figure 3 shows a 10 μm x 10 μm picture of the topography (upper-left image) of one of the magnetic graphite samples, and the corresponding magnetic behavior (MFM, upper-right image). A large pore can be seen at the topography image, and a grid scheme typical of magnetic systems is clearly seen surrounding it, where magnetic domains are fairly homogeneously distributed. The periodicity of the domains is approximately 1 μm. At the bottom of those figures we show the corresponding 3D-images. These results led us to conclude that the magnetic signal revealed by the phase shifts, is not correlated to the roughness of the sample surface outside the pore, i.e., the origin of the magnetic grid seems to be genuinely intrinsic.

Figure 4a. shows the magnetic behavior of the reported material, for an external magnetic field of 1000 Oe. We compare the magnetic behavior between the pristine and the modified magnetic graphite. The magnetic behavior of the pristine sample is the expected one, i.e. the well-known diamagnetic response of graphite. We did not subtract any background in this curve, which demonstrates the intensity of the ferromagnetic signal of the sample. The inset shows plots of the product $\chi T$, as a function of temperature, where $\chi = m/H$. The curve with open symbols was taken after one week of sample preparation and the curve with closed symbols was taken after six weeks. A ferromagnetic transition is observed around 125 K in both curves, attesting the temporal stability of the magnetic graphite here reported. The second measurement was done up to 350 K, revealing another ferromagnetic transition around 320 K (see arrows). This complex behavior can be explained through the inhomogeneity of the modified magnetic carbon specimen, which could not be avoided. As a consequence of the defects created by the procedure here reported, the material becomes inhomogeneous, which seems to be an essential feature of the magnetic graphite samples. Due to the defects created the material enhances the intrinsic magnetic behavior previously reported[2,3]. The use of powder graphite as a reagent seems to be crucial because of the high reactivity to chemical attacks as the one here described, due to the high exposed surface and the previous existence of defects. Thus, the concentration of defects produced by chemical attack, could be the cause to produce the strong magnetic response here reported.

Figure 4b. exhibits the hysteresis cycle for the m vs. H curves at 4.2, 200 and 300 K. We assumed that the material at 4.2 K lacked of paramagnetic phases and used this curve to establish the diamagnetic contribution to be subtracted at all temperatures. Then, a paramagnetic contribution was also subtracted at 200 and 300 K. Figure 4b. shows the curves after these subtractions -in pairs, to improve clarity- and the upper left inset exhibits the 200 K hysteresis cycle before any background was subtracted. The saturation magnetic moment was very strong, 0.42 emu/g, 0.28 emu/g and 0.25 emu/g, for T = 4.2, 200 and 300

K, respectively. To justify these values through the role of magnetic impurities by assuming that all these impurities behave as bulk ferromagnetic material -which most probably would not be the case-, it would be required about 1900 ppm of Fe. Since this value is much higher than the one determined as the total content of Fe in the studied samples, this justification could be ruled out. A decrease in the coercive field was observed in the 200-300 K region (Hc = 850 and 350 Oe, at 200 and 300 K, respectively). The remnant magnetization was 0.093, 0.075 and 0.04 emu/g, which corresponded to 22, 26 and 16 % of the saturation magnetic moment for 4.2, 200 and 300 K, respectively.

In conclusion, we have found a simple and inexpensive chemical route, based on a vapor phase reaction, to obtain ferromagnetic graphite. According to theoretical studies[1], defects in the honeycomb structure of graphite could develop spontaneous magnetization due to the rise of a sharp asymmetric peak in the density of states at the Fermi level, which is required for a system of itinerant electrons to show ferromagnetism. The magnetic behavior here reported would be associated to microstructural features observed in the attacked sample that produce an inhomogeneous material with enhanced magnetism. The $sp^3$ and $sp^2$ bonds could play a role that should be investigated in the future.

**Acknowledgements**

The authors wish to thank PEDECIBA and CSIC (Uruguayan Organizations) and CNPq and FAPESP (Brazilian Organizations) for financial support. We gratefully acknowledge E. Longo, I. G. Gobato, A. Vercik, J. A. Chiquito, E. Marega, A. V. Narlikar, P. F. S. Moraes (*in memoriam*) and J. C. Ortega for valuable discussions and help in the initial steps of this work. We also thank the technical help of A. Márquez (SEM), A. Altamirano (AAS) A. Sixto and P. Noblía (XRF), R. L. Almeida (magnetization measurements), and F. C. Rangel (MFM).

**Correspondence** and requests for materials should be addressed to A.W.M. (e-mail: amombru@fq.edu.uy)

**Patent registered**

**Figure 1**. SEM images of the morphology of **a)** the graphite used as reagent, **b)** a region where many pores of different sizes can be seen, **c)** one pore, 1 μm diameter –the propagation of the pore along the lamellar structure can be seen-, **d)** a lamellar stacking structure.

**Figure 2**. Raman spectra of the pristine and modified magnetic graphite samples.

**Figure 3**. Set of images of an area of a magnetic graphite samples of 10 μm x 10 μm, corresponding to the topography (AFM, upper-left image) and the magnetic behavior (MFM, upper-right image). At the bottom we show the corresponding 3D-images.

**Figure 4**. Magnetic characterization of the magnetic graphite.

**a.** m vs. T curve of the magnetic graphite material, for H=1000 Oe; comparison of the magnetic behavior between the pristine graphite sample and the magnetic graphite. The inset shows the product $\chi T$ as a function of T for the same magnetic graphite sample, measured after one week (open symbols) and six weeks (closed symbols) of sample preparation. Arrows show the transition temperatures.

**b.** Hysteresis curves, m vs. H, for the magnetic graphite material. Comparison between T = 4.2 K (open symbols) and 300 K (closed symbols). The lower right inset shows the comparison between T = 200 K (open symbols) and 300 K (closed symbols). The upper left inset shows the hysteresis curve at T = 200 K, before the diamagnetic background was subtracted.

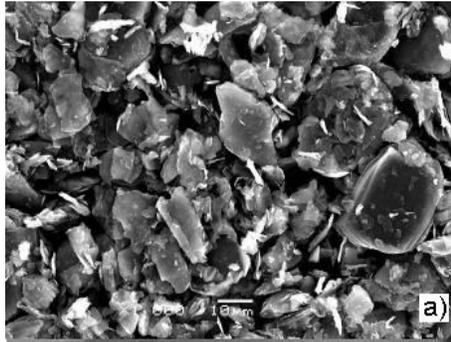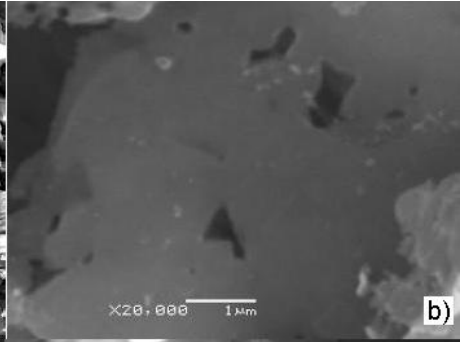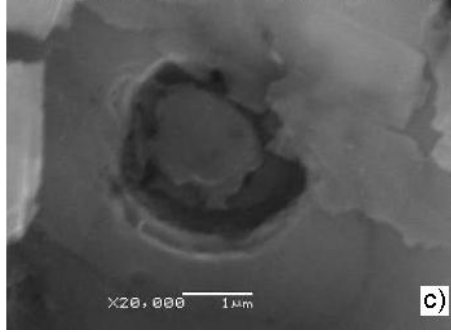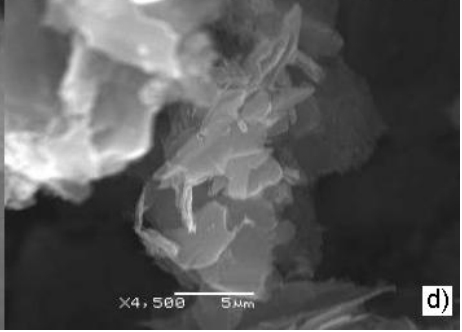

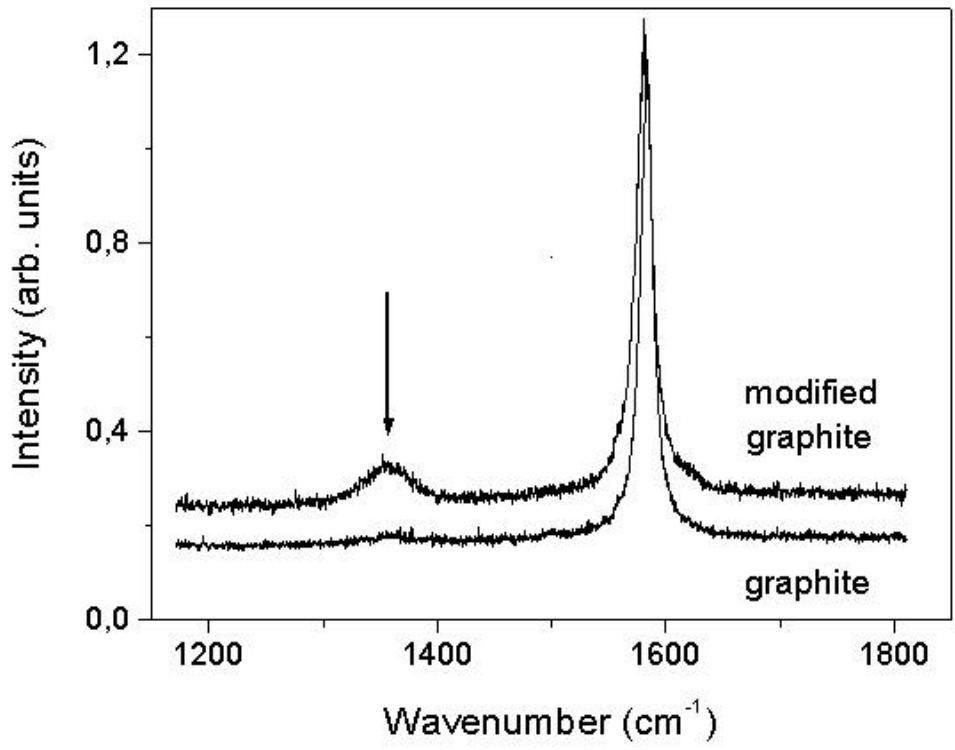

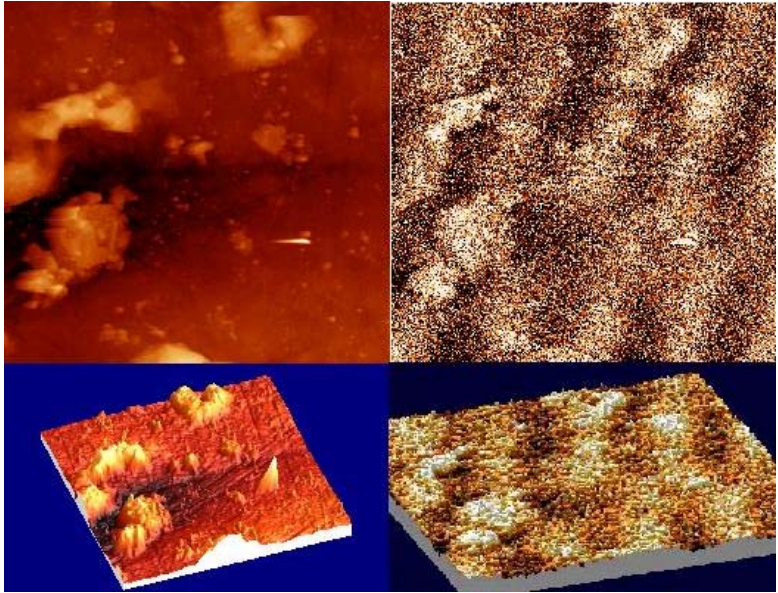

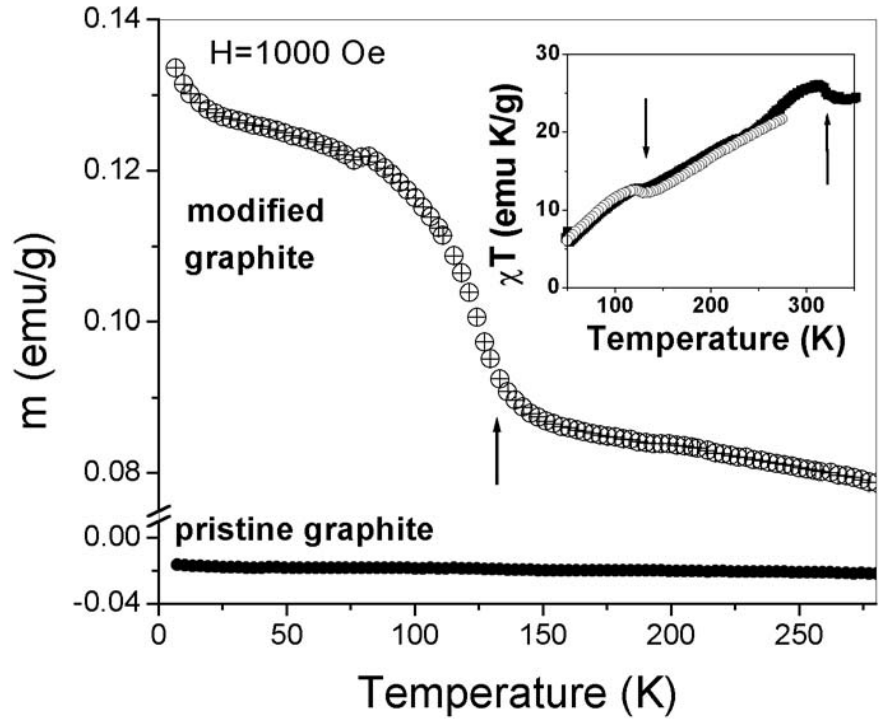

a)

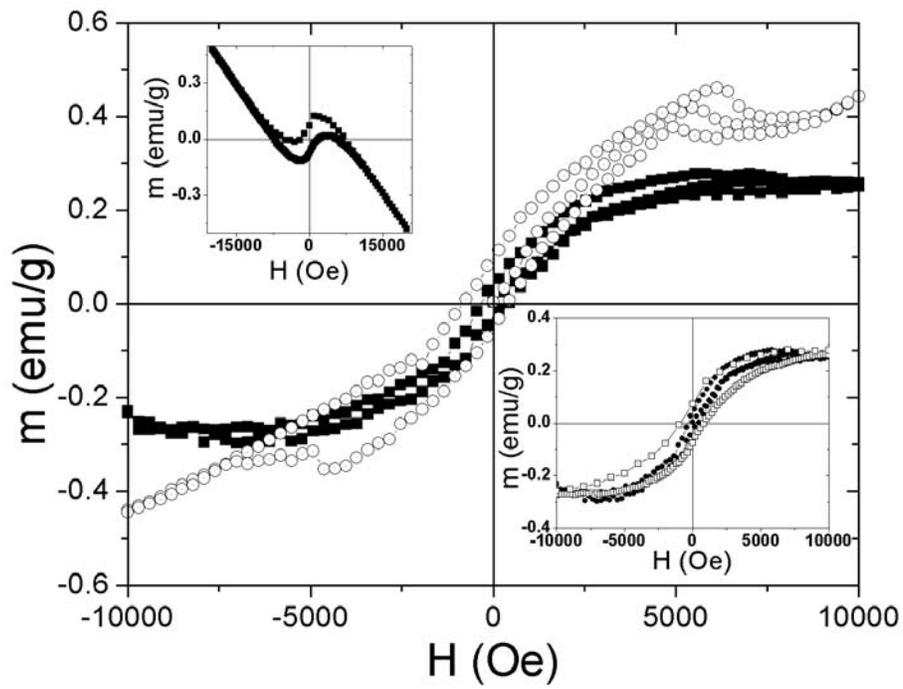

b)